\documentclass[aps,pra,amsfonts,amssymb,twocolumn,showpacs]{revtex4}
\usepackage{graphicx,hyperref,mathrsfs}
\def\be{\begin{equation}}
\def\ee{\end{equation}}
\def\ba{\begin{eqnarray}}
\def\ea{\end{eqnarray}}

\begin{document}
\title{Quasi-energies and Floquet states of two weakly coupled
 Bose-Einstein condensates
under periodic driving }

\author{Xiaobing Luo }
\affiliation{Institute of Physics, Chinese Academy of Sciences,
Beijing 100080, China}
\author{Qiongtao Xie }
\affiliation{Institute of Physics, Chinese Academy of Sciences,
Beijing 100080, China}
\author{Biao Wu }
\affiliation{Institute of Physics, Chinese Academy of Sciences,
Beijing 100080, China}

\begin{abstract}
We investigate the quasi-energies and Floquet states of two weakly
coupled Bose-Einstein condensates driven by a periodic force. The
quasi-energies and Floquet states of this system are computed within
two different theoretical frameworks: the mean-field model and the
second-quantized model. The mean-field approach reveals a triangular
structure in the quasi-energy band. Our analysis of the
corresponding Floquet states shows that this triangle signals the
onset of a localization phenomenon, which can be regarded as a
generalization of the well-known phenomenon called coherent
destruction of tunneling. With the second quantized model, we find
also a triangular structure in the quantum quasi-energy band,
 which is
enveloped by the mean-field triangle. The close relation between
these two sets of quasi-energies is further explored by a
 semi-classical
method. With a Sommerfeld rule generalized to time-dependent
systems, the quantum quasi-energies are computed by quantizing
semiclassically the mean-field model and they are found to agree
very well with the
 results
obtained directly with the second-quantized model.

\end{abstract}

\maketitle

\section{Introduction}
Due to its simplicity, a single particle in a double-well potential
has been a paradigm to demonstrate many fundamental quantum
phenomena,  in particular, quantum tunneling and its
control\cite{P.Hanggi}. Immediately after the experimental creation
of Bose-Einstein condensates (BECs)  with dilute
 alkali
atomic gases\cite{Dalfovo, Leggett}, people realize the new
possibility of putting a BEC in a double-well potential and using it
to mimic this paradigm system to demonstrate experimentally quantum
tunneling and other fundamental quantum phenomena.  The subsequent
studies show that a BEC in a double-well potential has richer
physics due to interaction. For example, it was found that the
tunneling of BEC between the wells can be suppressed and therefore
self-trapped in one of the wells\cite{Smerzi}-\cite{Anderlini}. This
self-trapping phenomenon has now been observed experimentally with a
BEC\cite{Albiez,Levy}. More interestingly, the nonlinear two-mode
model derived to describe a BEC in a double-well potential was found
to be able to describe the tunneling between Bloch bands for a BEC
in an optical lattice\cite{nlz}. Due to interaction, a new quantum
phenomenon called nonlinear Landau-Zener tunneling was predicted and
later observed in experiment\cite{nlz,Cristiani}.

It is known that, for a single particle in a double-well potential,
one can use an external periodically driving field to control
quantum tunneling, either enhancing\cite{Lin}-\cite{Vorobeichik1} or
suppressing it\cite{Grossmann}-\cite{Jivulescu}  One then wonders
whether this kind of control can be also achieved for a BEC in a
double-well potential.  There have been several studies in this
 regard\cite{Holthaus}-\cite{Abdullaev}.  These studies indeed find
 that the periodically driving force can strongly affect the tunneling
between two weakly coupled BECs and therefore be used to control the
tunneling. Recently, we found that such a control of quantum
tunneling can also be achieved in an optical waveguide
system\cite{Luo} and be
 used to improve the
performance of an all-optical switch\cite{wub}.

In this paper we investigate the quasi-energies and Floquet states
of two weakly coupled BECs under periodic driving, which can be
realized experimentally with either a double-well potential or an
optical lattice\cite{Morsch}. Quasi-energies and Floquet states are
two
 basic
concepts and tools in describing and understanding periodically
driving systems. One can use either a mean-field nonlinear two-mode
model or a second quantized model to describe such a system. In this
paper we use both models to compute the quasi-energies and Floquet
states. In the mean-field two-mode model, we discover that there can
be more than two Floquet states and quasi-energies in a certain
range of parameters that characterize the amplitude and frequency of
the modulating force. With these additional Floquet states, there
appears a triangle in the quasi-energy levels. This triangular
structure in quasi-energies turns out to be crucial to understanding
the
 localization
phenomenon that has been found and studied
 previously\cite{Holthaus,Wang,Tsukada}.
Our analysis shows that the localization phenomenon can be regarded
as a generalization of a well-known phenomenon called coherent
destruction of tunneling(CDT). Therefore, we call it nonlinear
coherent destruction of tunneling(NCDT)\cite{Luo}.

In the second quantized model, our computation also reveals a
triangular structure in the quasi-energy levels. Interestingly, the
quantum triangle is enveloped perfectly by the mean-field triangle,
indicating a close connection between these two different
approaches. By analyzing the corresponding Floquet states, we find
that this quantum triangle of quasi-energies is also connected to
the localization phenomenon called NCDT. The close relation between
quantum quasi-energies and mean-field quasi-energies is further
explored by a semi-classical method. By using a Sommerfeld
quantization rule adapted for a time-dependent system, we
re-calculate the quantum quasi-energies by quantizing
semiclassically the mean-field model. The results match very well
with the quantum quasi-energies obtained by directly using the
second quantized model.

Due to the complication brought by the chaos in the region of
moderate frequencies, the focus of our paper is on cases of high
frequency modulation.


\section{Quasi-energies and Floquet states}
We consider a system of $N$ identical bosons, which can occupy only
two quantum states. If there is interaction between bosons, the
system Hamiltonian reads\cite{Leggett}
\begin{eqnarray}
H_{q}&=&\frac{\gamma}{2}(\hat{a}^{\dag}\hat{a}-\hat{b}^{\dag}\hat{b})
-\frac{v}{2}(\hat{a}^{\dag}\hat{b}+\hat{a}\hat{b}^{\dag})\nonumber \\
&&+\frac{c}{2N}(\hat{a}^{\dag}\hat{a}^{\dag}\hat{}a\hat{a}+
\hat{b}^{\dag}\hat{b}^{\dag}\hat{b}\hat{b})\,, \label{eq:hqfock}
\end{eqnarray}
where $\gamma$ is the energy difference between the two quantum
states denoted by $\hat{a}^\dag,\hat{a}$ and $\hat{b}^\dag,\hat{b}$
and $v$ is
 the
coupling constant between the two modes. The interaction strength is
 given by
\be c=\frac{4\pi\hbar^2 a_s}{m}\int |\psi_0(\vec{r})|^4d\vec{r}\,,
\ee where we have used a reasonable assumption that the wave
functions of the two quantum states are the same except a possible
trivial shift of the center and the wave function is normalized
$\int
 |\psi_0(\vec{r})|^2d\vec{r}=1$.

When the temperature is very low so that we can ignore any thermal
 effect and
at the same time the number of bosons $N$ is very large, it is
 appropriate
to make the following coherent substitutes \be a=\langle
\hat{a}\rangle/\sqrt{N}\,,~~~~~~b=\langle
 \hat{b}\rangle/\sqrt{N}\,.
\ee This leads to a mean-field Hamiltonian \ba
H_{mf}&=&\frac{\langle H_q\rangle}{N}=
\frac{\gamma}{2}(|a|^2-|b|^2)-\frac{v}{2}(a^{*}b+ab^{*})\nonumber\\
&&+\frac{c}{2}(|a|^4+|b|^4)\,. \label{eq:hm} \ea

The system described above has now been realized with a double-well
potential.  For the experiment in Ref.\cite{Albiez}, there are about
1150 atoms and a simple estimate gives $v\approx 65.3$s$^{-1}$ and
$c/v\approx 15$. This system can also be realized experimentally
with an optical lattice\cite{nlz,Morsch}.

In our study, we have $\gamma=A\cos(\omega t)$, that is, the energy
difference between the two quantum states is changed periodically.
With the double-well potential, this can be achieved by shifting
periodically the power of lasers that generate the double wells. For
an optical lattice, this can be accomplished by shaking along the
lattice direction. We focus our study on the quasi-energies and
Floquet states of this system as these are two basic concepts and
tools in understanding a periodically driving system.

\subsection{Mean-field model}
We first consider the mean-field model. From the mean-field
Hamiltonian (\ref{eq:hm}), we can obtain a two-mode Gross-Pitaevskii
equation
\begin{eqnarray}
i\frac{d}{dt}\left(\matrix{a \cr b} \right)=
\left(\matrix{\frac{\gamma}{2}+c|a|^2 & -\frac{v}{2} \cr
-\frac{v}{2} & -\frac{\gamma}{2}+c|b|^2} \right ) \left(\matrix{a
\cr b} \right ), \label{eq:meanfield}
\end{eqnarray}
where we have used the natural unit $\hbar=1$. Although the
parameters $c$, $v$, $A$, and $\omega$ are of unit of energy, we
shall treat them as dimensionless parameters in the following
discussion because what is essential is the ratios between these
parameters not their absolute values.

Like its linear counterpart, a nonlinear periodic time dependent
equation admits solutions in the form of Floquet states. For
Eq.(\ref{eq:meanfield}), its Floquet state has the following form
\be \left(\matrix{a \cr b} \right)=e^{-i\varepsilon t}
\left(\matrix{\phi_1(t) \cr \phi_2(t)} \right)\,, \ee where both
$\phi_1(t)$ and $\phi_2(t)$ are periodic functions of period of
$T=2\pi/\omega$ and the constant $\varepsilon$ is the corresponding
quasi-energy. After one period, this solution returns to its
original state by picking up an extra phase of $\varepsilon T$. To
calculate numerically Floquet states and quasi-energies, we follow
the strategy that was used to compute nonlinear Bloch states and the
eigen-energies\cite{NJP}. In this strategy, we expand the Floquet
states in Fourier series \be
\phi_{1}=\sum_{n=-L}^{L}a_{n}e^{in\omega t},~~~
\phi_{2}=\sum_{n=-L}^{L}b_{n}e^{in\omega t}, \ee where $L$ is the
cut-off and equal to 10 in our computation. With the substitution of
the above Fourier series into Eq. (\ref{eq:meanfield}), one can
obtain $4L+2$ equalities for the coefficients of each Fourier term
$e^{in\omega t}$.  The Floquet state and the quasi-energy are found
by finding the roots of this set of $4L+2$ nonlinear equations. Our
method is different from the previous methods used to compute
Floquet states and quasi-energies\cite{Holthaus}. We believe that it
is more powerful. For example, it can find the Floquet states that
correspond to hyperbolic fixed points in Poincar\'e section, which
can not be found with the previous method\cite{Holthaus}.

\begin{figure}[htp]
\center
\includegraphics[width=7.0cm]{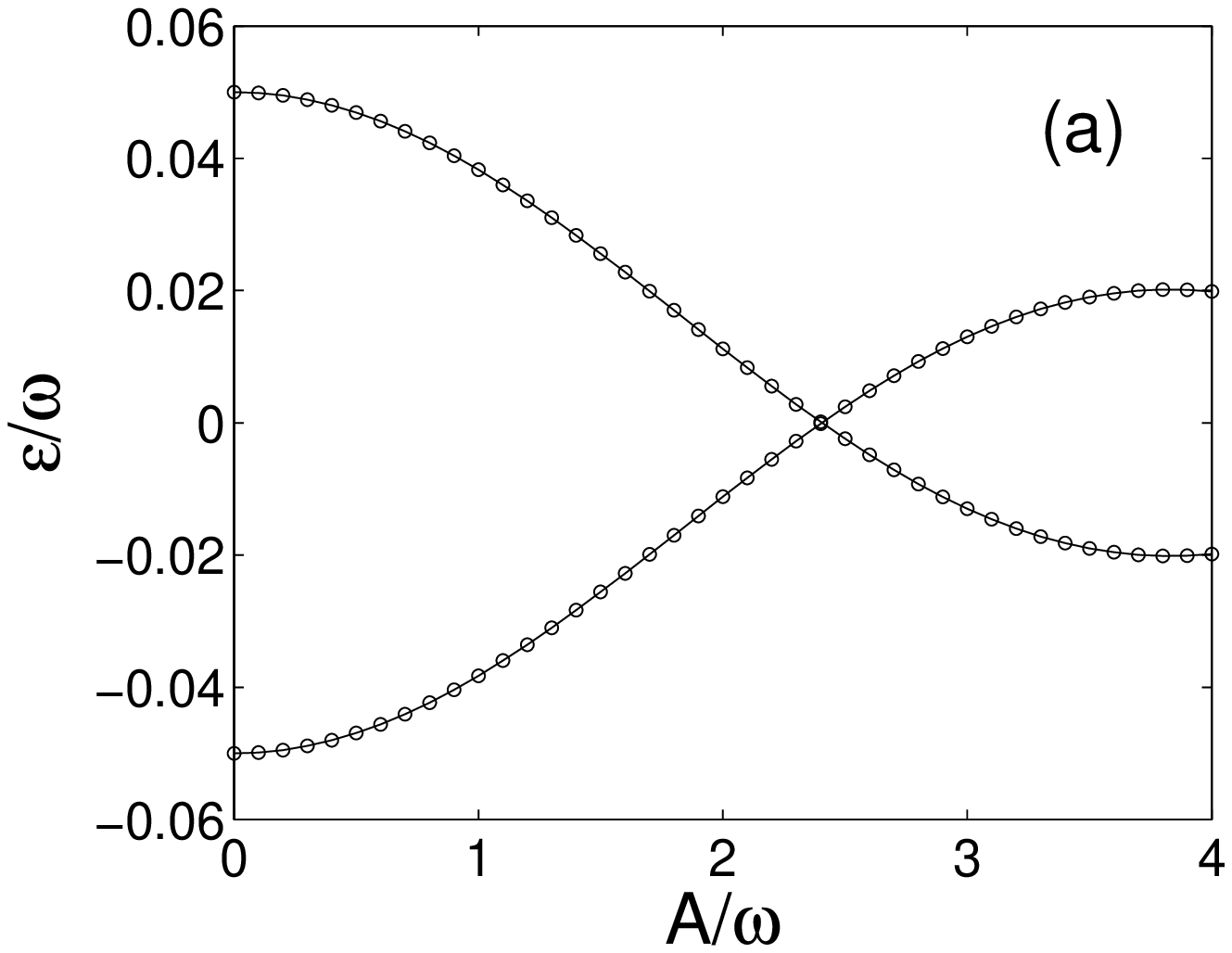}
\includegraphics[width=7.0cm]{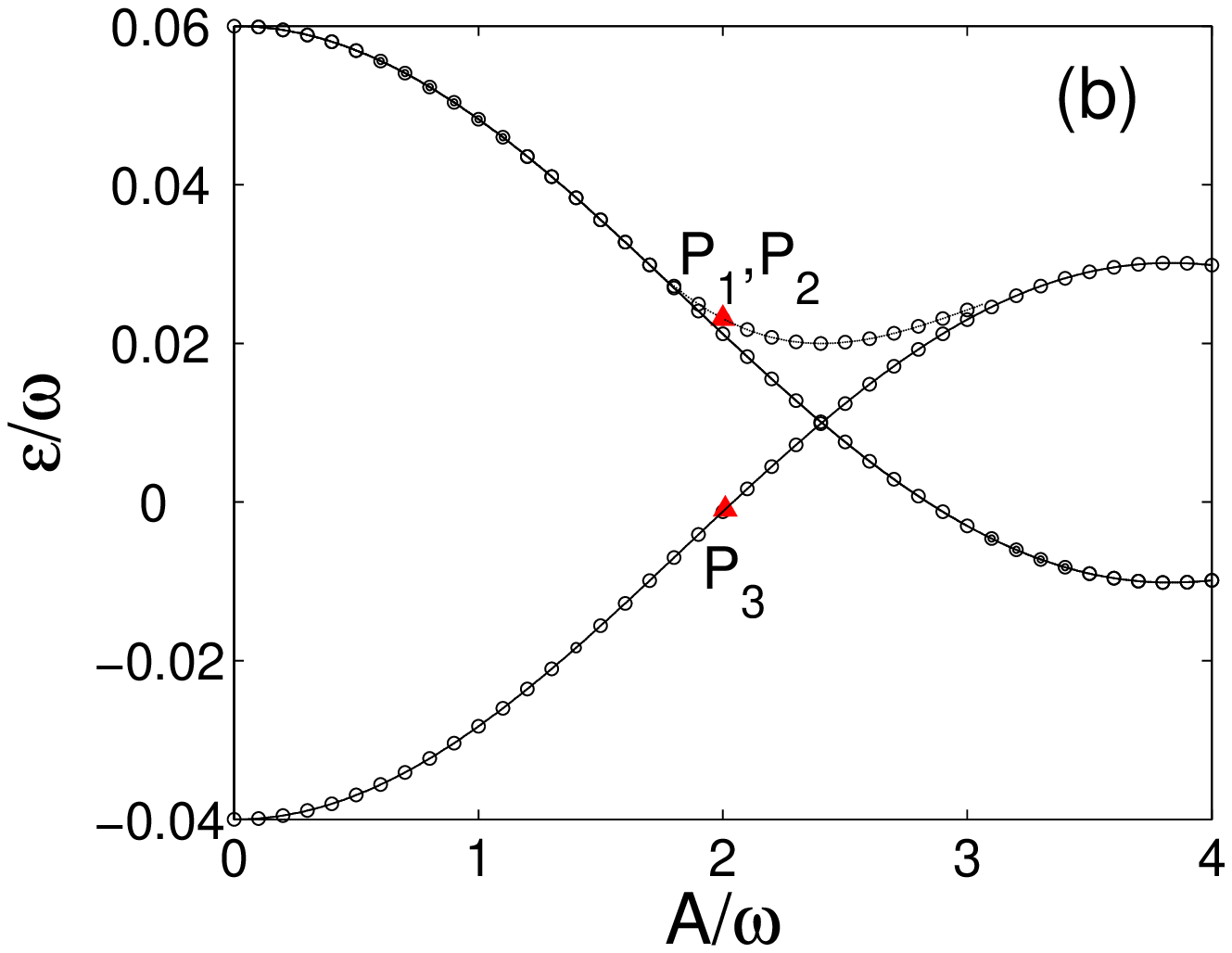}
\includegraphics[width=7.0cm]{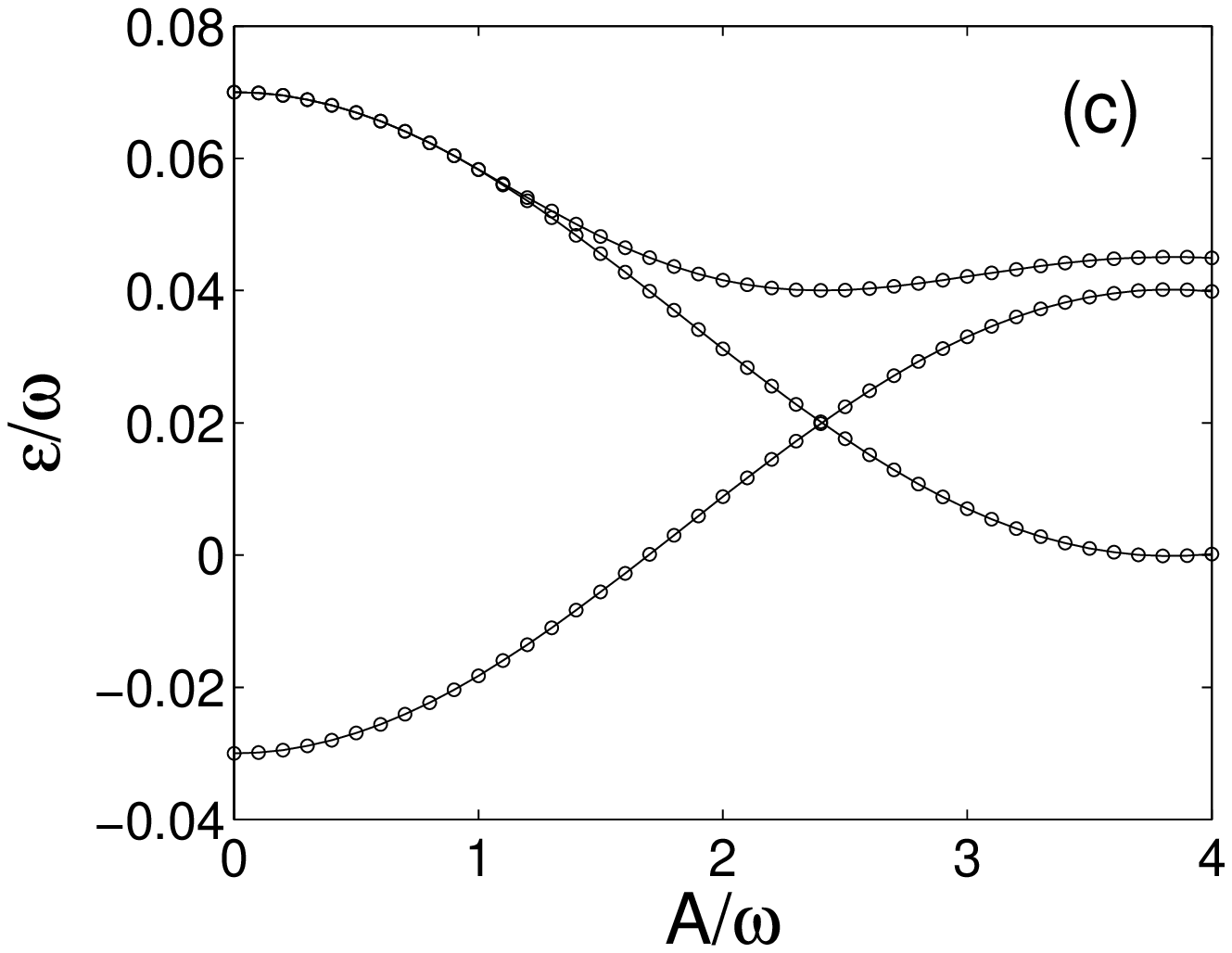}
\caption{(Color online)Quasi-energies as a function of $A/\omega$ at
(a)$c=0$; (b)$c=0.4$; (c)$c=0.8$. Solid lines are numerical results
and circles are the approximate analytical results for high
frequencies with Eqs.(\ref{highfrequency1},\ref{highfrequency2}).
$v=1,\omega=10$.}\label{fig:meanfieldquasienergies}
\end{figure}

Our numerical results of quasi-energies are plotted in
 Fig.\ref{fig:meanfieldquasienergies}.
It is clear from Fig.\ref{fig:meanfieldquasienergies} that, for the
linear case, there are two quasi-energies at a given value of
$A/\omega$ with one isolated degeneracy point. For the nonlinear
case, we notice that there are three quasi-energies within a certain
range of $A/\omega$ with two of them degenerate. The three
quasi-energies form a triangle in the quasi-energy levels as seen in
Fig.\ref{fig:meanfieldquasienergies}(b). Among the three
quasi-energies, two quasi-energy levels are similar to their linear
counterparts with one isolated degenerate point while the third
quasi-energy level has no linear counterpart. Moreover, the third
quasi-energy is degenerate and corresponds to two different Floquet
states; this is indicated by marking the same point in
 Fig.\ref{fig:meanfieldquasienergies}(b)
with two symbols $P_1$ and $P_2$. Note two things: (1) there is no
threshold value of $c$ for the triangle to appear; (2) the right
corner of triangle is open for relatively larger nonlinear parameter
$c$.

Despite the obvious similarity between the nonlinear Floquet states
and the linear ones, there are a couple of conceptual differences.
(1) A periodically driven $n$-level linear system possesses
precisely
 $n$
Floquet states whereas the number of nonlinear Floquet states of
$n$-mode system can be bigger than $n$ as we have witnessed above.
(2) In linear case, all wave functions can be decomposed into a
superposition of Floquet states and, therefore, the dynamics of the
system is dictated by Floquet states. In the nonlinear case, the
superposition principle breaks down, the dynamics of the system can
no longer be completely determined by Floquet states.

The triangular structure of the quasi-energy is very similar to the
energy loop discovered within the context of nonlinear Landau-Zener
tunneling\cite{nlz}. In fact, they are mathematically related. For
high frequencies, $\omega\gg \max\{v,c\}$, we take advantage of the
transformation \ba a=a' \exp[-i\frac{A\sin(\omega t)}{2\omega}],~~
b=b^\prime\exp[i\frac{A\sin(\omega t)}{2\omega}]. \label{trans} \ea
After averaging out the high frequency terms\cite{Wang,Kayanuma}, we
obtain a non-driving nonlinear model,
\begin{eqnarray}
i\dot{a'}&=&-\frac{v}{2}J_0(A/\omega)b'+c|a'|^{2} a',
\label{highfrequency1}\\
i\dot{b'}&=&-\frac{v}{2}J_0(A/\omega)a'+c|b'|^{2}
b'\label{highfrequency2},
\end{eqnarray}
where $J_0$ is the zeroth-order Bessel function.  It is clear from
the transformation in Eq.(\ref{trans}) that the eigenstates of the
above non-driving nonlinear equations correspond to the Floquet
states of Eq.(\ref{eq:meanfield}). We have computed the eigenstates
of Eqs.(\ref{highfrequency1},\ref{highfrequency2}) and the
corresponding eigenenergies, which are plotted as circles in
Fig.\ref{fig:meanfieldquasienergies}. The consistency with our
previous numerical results is obvious. As is known in
Ref.\cite{nlz},
 the
above nonlinear model admits additional eigenstates when
$c>J_0(A/\omega)v$. Therefore, this can be regarded as the condition
for the extra Floquet states to appear for the driving nonlinear
model Eq.\ref{eq:meanfield} at high frequencies. Since the Bessel
function $J_0(A/\omega)$ can be zero, there is no threshold value of
$c$ for the triangle to appear in the quasi-energy band.

\begin{figure}[htp]
\center
\includegraphics[width=8cm]{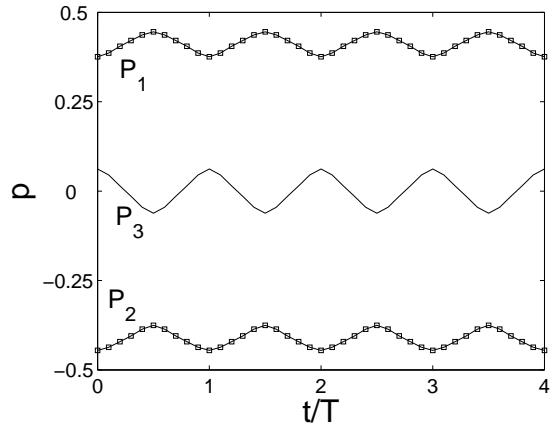}
\caption{Population imbalance $p$ for three stable nonlinear Floquet
states at $c=0.4, v=1, \omega=10, A/\omega=2.0$, for an interval of
four periods of the driving force(solid lines). The Floquet states
correspond to the quasi-energies in
 Fig.\ref{fig:meanfieldquasienergies}(b)
marked as $P_1, P_2, P_3$ with triangles. The squares are for
Floquet states in the highest two quantum quasi-energy levels with
$N=500$.} \label{fig:floquetstate}
\end{figure}

The nonlinear Floquet states are also examined thoroughly. We find
that some of them are localized, which is very different from the
linear Floquet states that are always unlocalized. To describe
localization, we introduce a new variable, $p=(|a|^2-|b|^2)/2$,
which measures the population difference between the two modes. One
Floquet state is localized if the average of $p$ over one period,
\be \langle p\rangle_t=\frac{1}{T}\int_{0}^{T}dt\,p(t)\,,
\label{eq:local} \ee is nonzero; it is unlocalized if  $\langle
p\rangle_t=0$. In Fig.\ref{fig:floquetstate} , the population
difference $p$ is plotted as a function of time for three stable
nonlinear Floquet states marked as $P_1, P_2, P_3$ in
Fig.\ref{fig:meanfieldquasienergies}(b). Evidently, one of these
states is unlocalized since $p$ oscillates around zero. However, two
other states are localized with $p$ oscillating around a non-zero
value. The localization means that the BEC described by such Floquet
states tends to stay in one mode and reluctant to tunnel to the
other mode. Therefore, localization can be understood as a
suppression of tunneling. Our study shows that on one hand, all the
localized Floquet states correspond to the highest
 quasi-energies on the
triangle and on the other hand, all Floquet states in the linear
case
 and all the Floquet
states not related to the quasi-energy triangle are not localized.
This implies that the triangle in
Fig.\ref{fig:meanfieldquasienergies} is related to localization or
suppression of tunneling. This is indeed the case as we have shown
in Ref.\cite{Luo}. We shall not repeat what we have done in
Ref.\cite{Luo}; we shall look into this connection from a different
angle.

\begin{figure}[htp]
\center
\includegraphics[width=9cm]{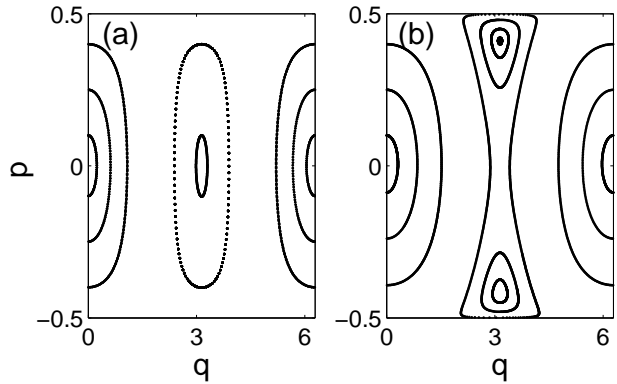}
\caption{Poincar\'e surface of section of the Hamiltonian
(\ref{eq:classical}). (a)$A/\omega=0.1$; (b)$A/\omega=2.0$. Other
parameters are $c=0.4$, $v=1$, $\omega=10$.} \label{fig:phase}
\end{figure}

In the mean-field model (\ref{eq:meanfield}), the norm
$|a|^2+|b|^2=1$ is conserved and the overall phase is not essential
to the dynamics. Therefore, we can reduce the complex dynamical
variables
 $a=|a|e^{i\theta_{a}}$,
$b=|b|e^{i\theta_{b}}$ to a pair of real variable,
$p=(|a|^2-|b|^2)/2$
 and
the relative phase $q=\theta_{b}-\theta_{a}$. In terms of $p$ and
$q$, the mean-field Hamiltonian (\ref{eq:hm}) becomes
\begin{eqnarray}
H_{cl}=Ap\cos(\omega t)-\frac{v}{2}\sqrt{1-4p^2}\cos q+
\frac{c}{4}(4p^2+1). \label{eq:classical}
\end{eqnarray}
As $p$ and $q$ are canonically conjugate variables of the above
classical Hamiltonian system, one can derive a set of equations of
motion. From the equations of motion, one can plot the Poincar\'e
section of this system. Two Poincar\'e sections are illustrated in
Fig.\ref{fig:phase} for two sets of parameters. As the overall phase
is removed, the Floquet states correspond to the fixed points in
Poincar\'e section.

The parameters for Fig.\ref{fig:phase}(a) are outside the triangle
range. In this figure, there are only two fixed points located at
$p=0$ and all the motions around the fixed points are oscillating
around $p=0$, indicating no localization or suppression of
tunneling. The situation is different in Fig.\ref{fig:phase}(b),
whose parameters lie in the triangle range. In
Fig.\ref{fig:phase}(b), there are four fixed points: one at $q=0$(or
$2\pi$); three at $q=\pi$. Among the three at $q=\pi$, one is
hyperbolic and unstable whereas the other two are not only stable
but localized. Moreover, all the orbits surrounding these two stable
fixed points at $q=\pi$ are localized solutions. These again show
that the triangle structure in quasi-energies are related to
localization or suppression of tunneling.

In Ref.\cite{Luo}, the localization phenomenon discussed above is
called nonlinear coherent destruction of tunneling (NCDT). There are
two reasons for this. First, the degeneracy point in
Fig.\ref{fig:meanfieldquasienergies}(a) is related to a localization
phenomenon called coherent destruction of tunneling (CDT) and the
triangle can be seen as the result of enlargement of the degenerate
point by nonlinearity. Second, as we have seen in
Fig.\ref{fig:phase}, the localization phenomenon is intimately
related to the nonlinear Floquet states and we know that CDT is
related to linear Floquet states. The localization phenomenon which
we call NCDT has been called in literature self-trapping or, more
precisely, periodically
 modulated
self-trapping\cite{Holthaus,Wang,Tsukada}.

\subsection{Second quantized model}
We now turn to the second quantized model (\ref{eq:hqfock}) and
compute its Floquet states and quasi-energies. For a non-driving
system, it is well known that the eigen-energies and eigenstates of
the second quantized model are closely connected to its mean-field
counterparts\cite{Karkuszewski,Biao}. For this periodically driving
system, we want to explore how its quantum Floquet states and
quasi-energies are related to its mean-field counterparts and the
localization phenomenon called NCDT.

\begin{figure}[!tb]
\center
\includegraphics[width=8cm]{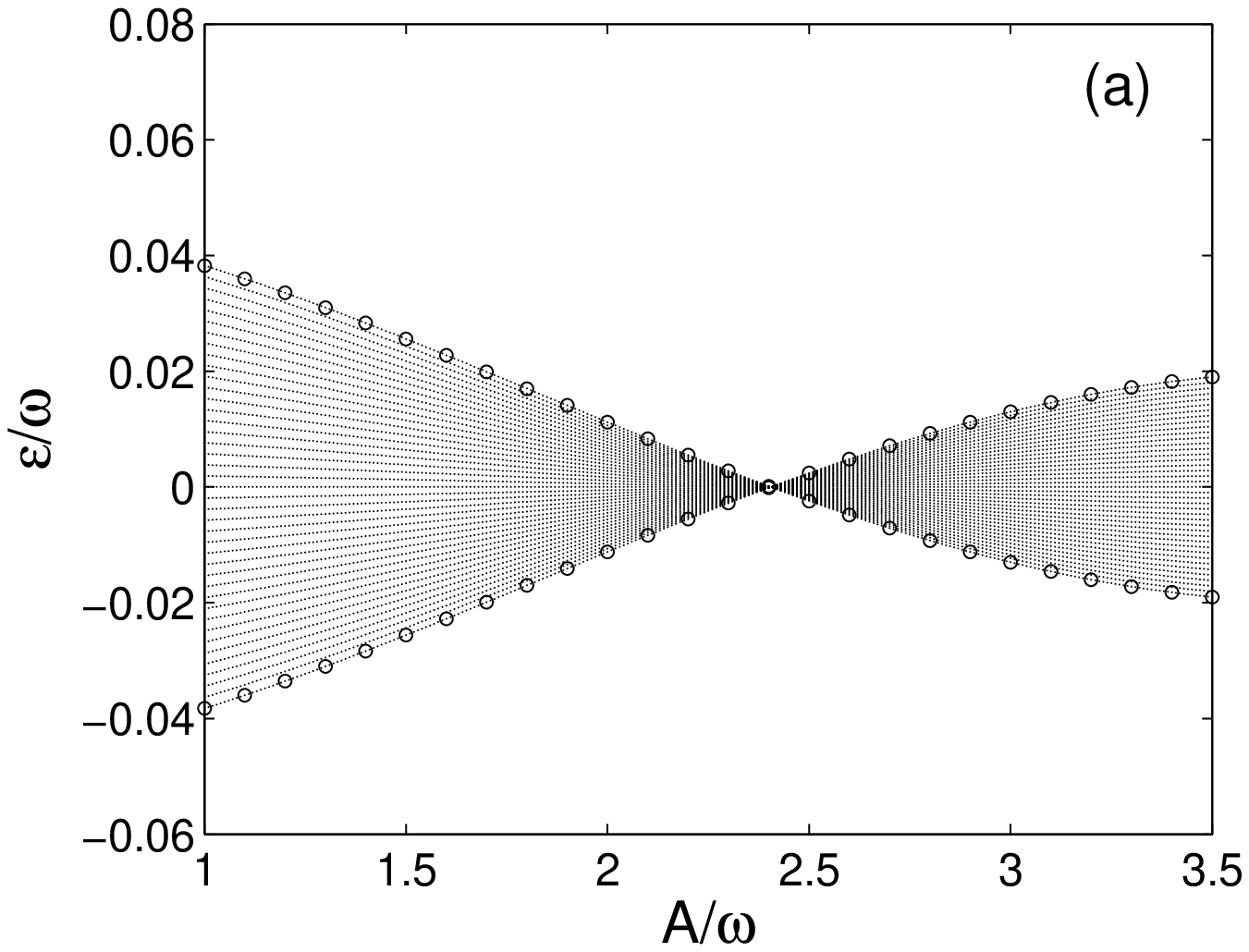}
\includegraphics[width=8cm]{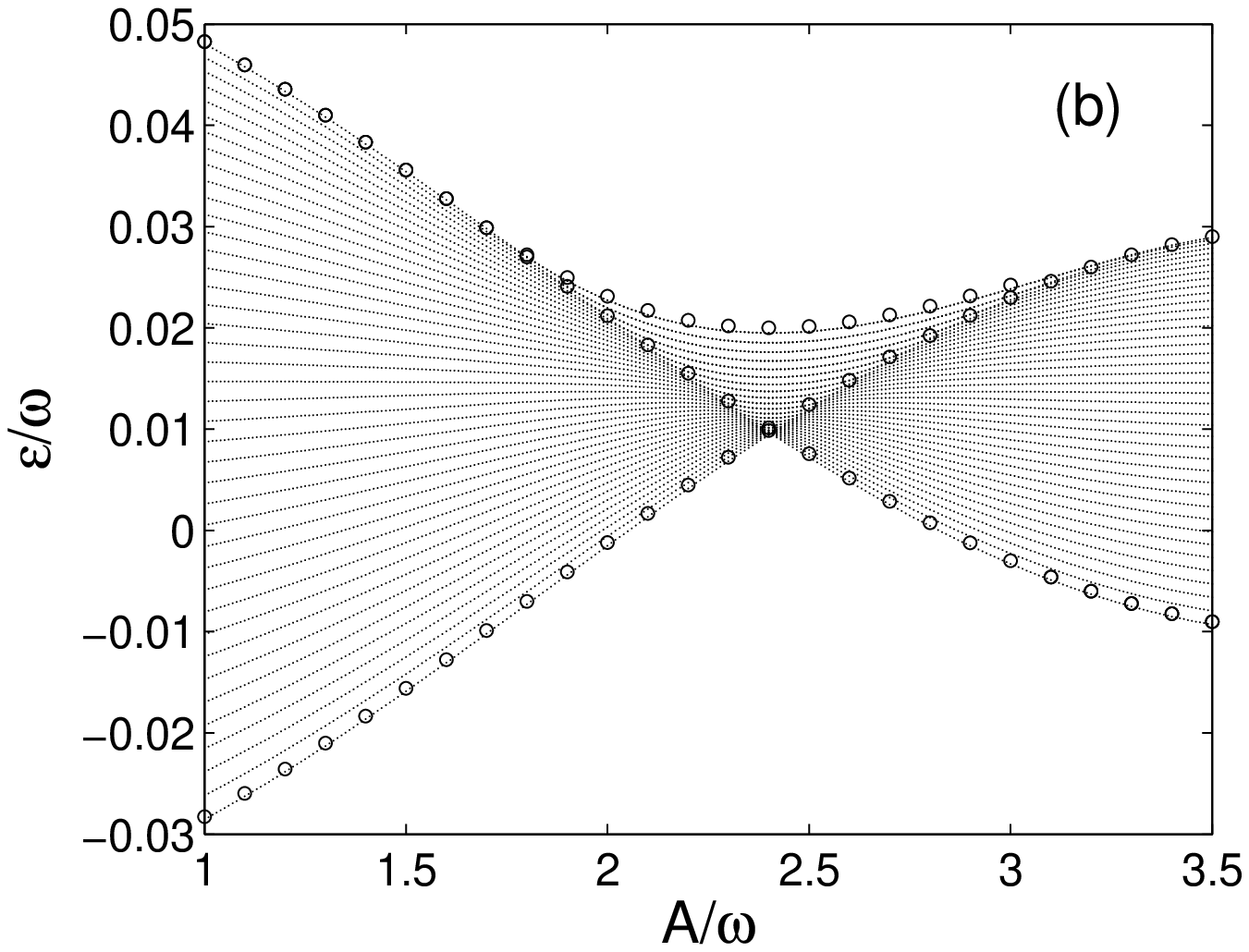}
\includegraphics[width=8cm]{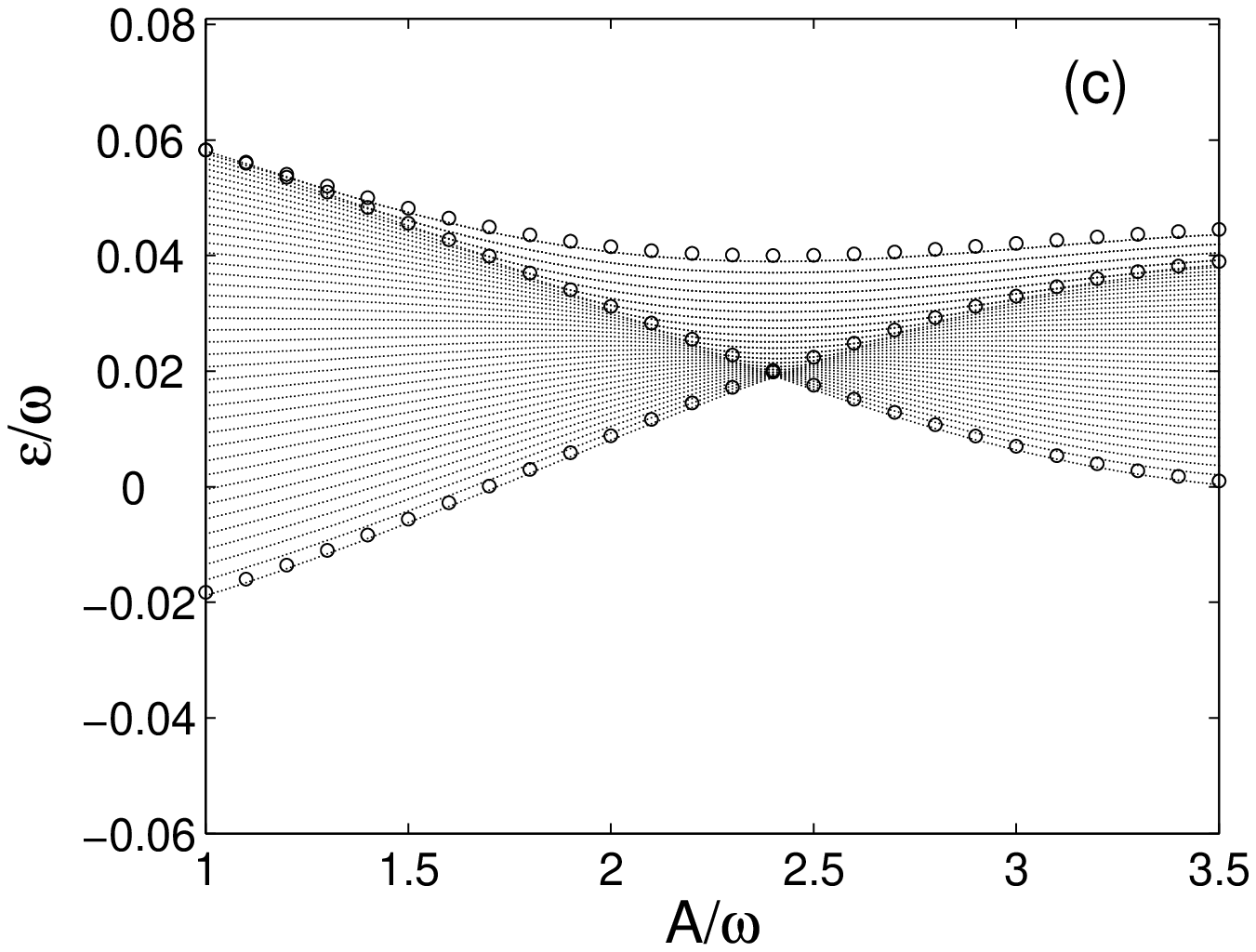}
\caption{Quantum quasi-energies ($N=40$) as a function of $A/\omega$
at $v=1, \omega=10$ for (a) $c=0.0$; (b)$c=0.4$; (c)$c=0.8$. The
open circles are mean-field quasi-energies. Note that for comparison
with mean-field theory, the quantum quasi-energies have been divided
by $N$.} \label{fig:correspondence}
\end{figure}

We follow the well established Floquet theory for a quantum
system\cite{sambe,shirley} to compute numerically quantum Floquet
states and quasi-energies. In the process, we have converted the
second quantized Hamiltonian (\ref{eq:hqfock}) into a pseudo-spin
Hamiltonian by introducing three angular momentum operators
 $\hat{J}_{x}=(\hat{a}^{\dag}\hat{b}+\hat{b}^{\dag}\hat{a})/2$,
$\hat{J}_{y}=i(\hat{b}^{\dag}\hat{a}-\hat{b}\hat{a}^{\dag})/2$, and
$\hat{J}_{z}=(\hat{a}^{\dag}\hat{a}-\hat{b}^{\dag}\hat{b})/2$, for
which the Casimir invariant is $\hat{J}^2=(N/2)(N/2+1)$. The second
quantized Hamilton of the system then becomes
\begin{eqnarray}
H_{q}=-v\hat{J}_{x}+\frac{c}{N}\hat{J}_{z}^{2}+A\cos(\omega
t)\hat{J}_{z}+\frac{c}{4}(N-2). \label{eq:hqj}
\end{eqnarray}
With this transformation, our system of $N$ identical bosons becomes
a spin system, whose Hilbert space is spanned by $N+1$ spin states
$|J=N/2, J_{z}=M\rangle$ with $M=-N/2,-N/2+1,\cdots,N/2$.

Our numerical results for quantum quasi-energies for $N=40$ are
shown in Fig.\ref{fig:correspondence}. We immediately notice that
these quantum quasi-energy levels have very similar structures to
their mean-field counterparts. For the non-interacting case in
Fig.\ref{fig:correspondence}(a), there is a single degeneracy point.
For interacting cases in Fig.\ref{fig:correspondence}(b)\&(c), there
are triangular structures just as in the mean-field model. For
comparison, the mean-field quasi-energies are plotted as open
circles in Fig.\ref{fig:correspondence}. To one's amazement or
expectation, the quantum quasi-energies are bounded by the
mean-field results perfectly. Another interesting feature in
Fig.\ref{fig:correspondence} is that all the quasi-energies in the
triangle area is doubly degenerate and this degeneracy immediately
breaks up outside the triangle. The feature is related the
localization phenomenon NCDT as we shall discuss next.

\begin{figure}[htp]
\center
\includegraphics[width=7cm]{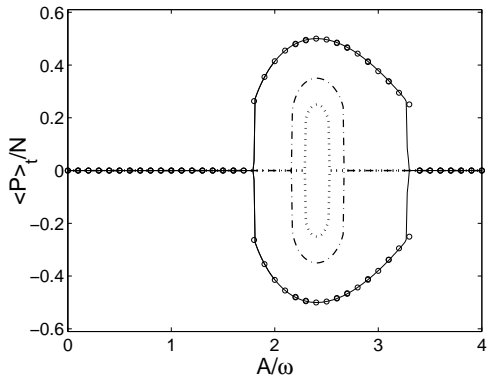}
\caption{Population difference $\langle P\rangle_t$ for every
Floquet state
 in
the highest two quantum quasi-energy levels(solid line), the 149th
and
 150th
quantum quasi-energy levels(dot-dashed line), and the 249th and
250th quantum quasi-energy levels(dotted line) at
$c/v=0.4,\omega/v=10, N=500$. The open circles are for the
population difference $\langle p\rangle_t$ for the highest
mean-field quasi-energy level in
Fig.\ref{fig:meanfieldquasienergies}(b).}
\label{fig:QuantumFloquetstate}
\end{figure}

There is also a close relation between quantum Floquet states and
mean-field Floquet states. We examine this relation in terms of
localization. To measure how a quantum Floquet state is localized,
we define \be \langle P\rangle_t=\frac{1}{T}\int_{0}^{T}dt\langle
u_n(t)|\hat{J}_z|u_n(t)\rangle \ee for a given Floquet state
$|u_n(t)\rangle$. This variable $\langle P\rangle_t$ quantifies the
population difference between the two modes. We have plotted this
variable for certain quantum Floquet states in Fig.
\ref{fig:QuantumFloquetstate}. It is apparent from this figure that
only the Floquet states for the quasi-energies inside the triangle
are localized. This again establishes the connection of the triangle
(quantum or mean-field) to the localization phenomenon NCDT. This
localization also explains why the Floquet states inside the
triangle are doubly degenerate. When localization occurs, there are
two equal possibilities. It can localize either in mode $a$ or in
mode $b$; this leads to degeneracy. The mean-field results are also
plotted in Fig.
 \ref{fig:QuantumFloquetstate}.
They match very well with the results for the two highest quantum
Floquet states. This good correspondence can be more clearly seen in
Fig. \ref{fig:floquetstate},  the temporal evolution of two highest
quantum Floquet sates agrees very well with the mean-field results
for an interval of four periods of the driving.

The quantum quasi-energies and Floquet state were studied in
 Ref.\cite{Holthaus}.
Their relation to the localization was also examined there. Our
primary purpose here is to compare them to the mean-field results
and explore their relations, which has not been studied so far.

\section{Semiclassical quantization}
In the previous section, we have demonstrated by direct numerical
computation how the quantum Floquet states and quasi-energies are
connected to their mean-field counterparts. This relation can be
further explored with a semiclassical method as the mean-field model
(\ref{eq:hm}) can be regarded as the classical limit of the second
quantized model (\ref{eq:hqfock}) in the limit of
$N\rightarrow\infty$\cite{Yaffe}. We shall follow the procedure in
Ref. \cite{Biao,Shankar,Garg,Graefe} and try to quantize the
classical Hamiltonian in Eq.(\ref{eq:classical}), which is
equivalent to Hamiltonian
 (\ref{eq:hm}),
with the Sommerfeld rule. However, as our system is time dependent,
the usual Sommerfeld quantization rule has to be generalized.

\begin{figure}[htp]
\center
\includegraphics[bb=25 212 640 563,width=8cm]{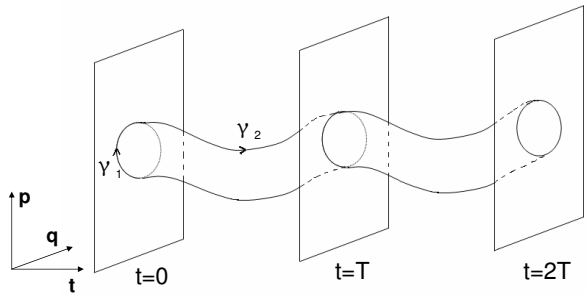}
\caption{Periodic vortex tube. Two paths are shown. The path
$\gamma_{1}$ lies in a plane of $t$=const and $\gamma_{2}$ is  a
path connecting a point $(p,q)$ at time $t$ with the same point at
time $t+T$.} \label{fig:tube}
\end{figure}

The generalization of the Sommerfeld rule has been done for any
time-dependent system\cite{Holthaus3,Korsch}. The basic idea is to
regard time as a dynamic variable and introduce a new canonical
momentum which conjugates time. We shall not go into the details of
this theory and shall only describe how this generalization works
for the case of our interest, a periodic time dependent system. As
seen in the Poincar\'e section of Fig.\ref{fig:phase}, there are
closed orbits around fixed points. These closed orbits will change
their positions and shapes in the phase space with time and return
to
 their
original points and shapes after one period. This kind of evolution
 forms a
tube in the space spanned by $p,q,t$ as depicted in
Fig.\ref{fig:tube}. This tube is called vortex tube. As the system
is periodic in time, the tube in Fig.\ref{fig:tube} is essentially a
torus.  The quantization can be done by choosing  two independent
closed paths on the vortex tube which cannot be homotopically
deformed onto each other and requiring
\begin{eqnarray}
\label{eq:i1}
I_{1}&=&\frac{1}{2\pi}\oint_{\gamma_{1}}pdq=n_{1}\hbar/N\,,\\
I_{2}&=&\frac{1}{2\pi}\oint_{\gamma_{2}}(pdq-H_{cl}dt)+
\frac{T}{2\pi}\varepsilon\nonumber\\
&=&n_{2}\hbar\,, \label{eq:i2}
\end{eqnarray}
where $n_1$ and $n_2$ non-negative integers. The quantization is
done in two steps: (1) we first find a path $\gamma_1$ that fulfills
the quantization condition for $I_{1}$; (2) the quantization
condition for $I_2$ is then used to compute the quasi-energy
 $\varepsilon$ as
\begin{eqnarray}
\varepsilon_{n_{1},n_{2}}=-\frac{1}{T}\oint_{\gamma_{2}}(pdq-H_{cl}dt)+n_{2}
\omega. \label{quasienergy}
\end{eqnarray}
In the above, $n_2\omega$ means that quasi-energy $\varepsilon$ is
only defined modulo $\omega$, reflecting the unique nature of
quasi-energy. One can view $\hbar/N$ in Eq.(\ref{eq:i1}) as the
effective Plank constant\cite{Biao,Shankar,Garg}, which goes to zero
at the limit of $N\rightarrow \infty$.

\begin{figure}[htp]
\center
\includegraphics[width=6cm]{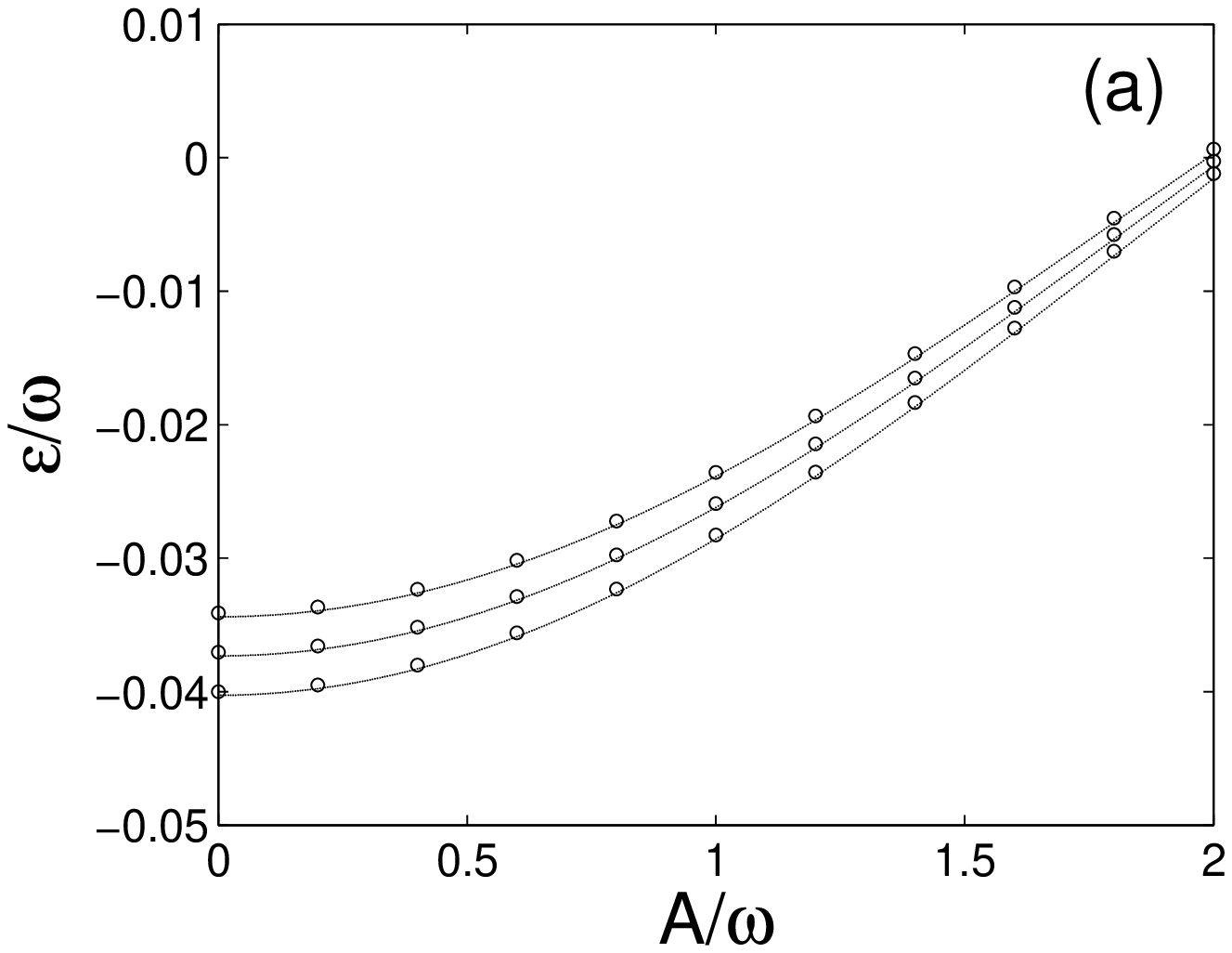}
\includegraphics[width=6cm]{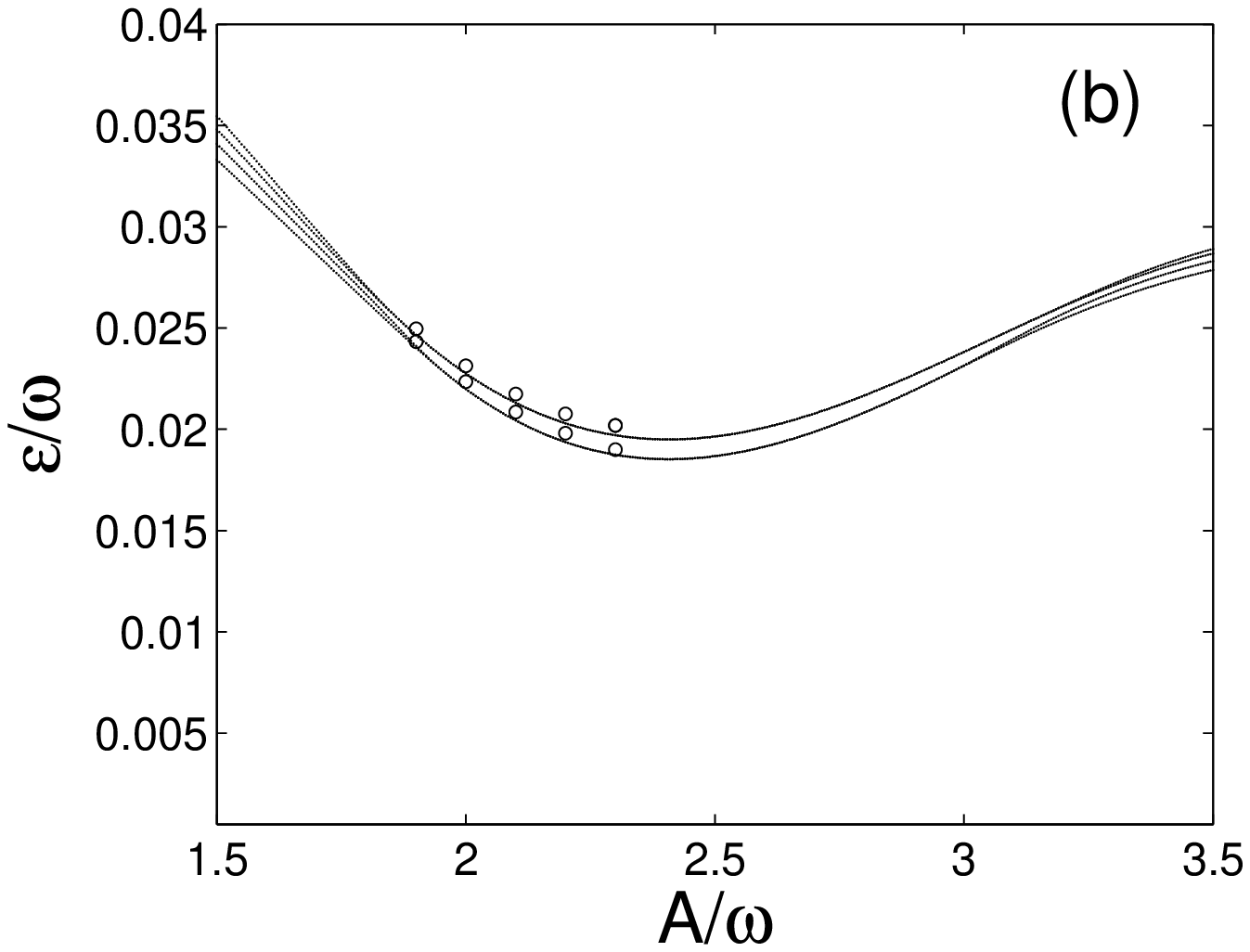}
\caption{Comparison between the quantum quasi-energy levels (solid
 lines)
with $N=40$ and semiclassical quasi-energy levels (open circles) at
$c=0.4, v=1,\omega=10$. (a) nondegenerate quasi-energy levels; (b)
 degenerate
quasi-energy levels. For clarity, we have only plotted a portion of
the quasi-energy levels.} \label{fig:semiclassic}
\end{figure}

Our semiclassical results of quasi-energies are plotted in
 Fig.\ref{fig:semiclassic}
to compare with the quantum quasi-energies obtained directly from
the second quantized model. They match perfectly, indicating the
success of the generalized Sommerfeld quantization rule. In our
 calculation,
the path $\gamma_1$ is chosen as the closed orbit in the Poincar\'e
section and $\gamma_2$ is the path along the maximal points of $p$
on the tube as illustrated in Fig.\ref{fig:tube}. Note that the
natural unit $\hbar=1$ is used in our calculation.

These semiclassical results are very helpful in understanding why
the quantum quasi-energies are enveloped by the mean-field
 quasi-energies
as seen in Fig.\ref{fig:correspondence}. We first look at the simple
case when there are only two fixed points in the Poincar\'e section,
as in Fig.\ref{fig:phase}(a). The fixed point at $q=0$ corresponds
to the nonlinear Floquet state with lower quasi-energy and the other
fixed point corresponds to the Floquet state with higher
quasi-energy. This implies that the quantization for orbits around
the fixed point at $q=0$ produces quasi-energies that are higher
than the corresponding mean-field quasi-energy and the quantization
for orbits around the fixed point at $q=\pi$ yields quasi-energies
that are lower than the corresponding mean-field quasi-energy. As a
result, the quantum quasi-energies are bounded by the mean-field
quasi-energies. The double degeneracy of the quantum quasi-energies
within the triangle can also be explained with this semiclassical
approach. As shown in Fig.\ref{fig:phase}(b), there are two stable
fixed points at $q=\pi$. These two fixed points correspond to two
Floquet states with the same quasi-energy. This indicates that if
one quantizes semiclassically the orbits around these two fixed
points, one would get two identical sets of quasi-energies. This
explains the double degeneracy.

\section{Conclusions}

To summarize, we have studied the quasi-energies and Floquet states
of two weakly coupled Bose-Einstein condensates subject to a
periodic driving. Both the mean-field model and the second quantized
model are used. A triangular structure was found in both mean-field
quasi-energy levels and quantum quasi-energy levels. Moreover, we
have revealed that the quantum quasi-energy levels are bound by
their mean-field counterparts and we have explained it with
semiclassical quantization. In addition, by looking into the Floquet
states, we have found that the triangle in the quasi-energies is
related a localization phenomenon which we call nonlinear coherent
destruction of tunneling (NCDT).

\acknowledgments{We thank the helpful discussion with Jie Liu. This
work is supported by the ``BaiRen'' program of Chinese Academy of
Sciences, the NSF of China (10504040), and the 973 project of
China(2005CB724500,2006CB921400).}

\end{document}